\documentstyle[epsfig,12pt]{article}
\begin{document}
\title{Formation of quantized vortices in
\\ a gaseous Bose-Einstein condensate}

\author{F. Chevy, K.~W. Madison, V. Bretin, and J. Dalibard\\
Laboratoire Kastler Brossel, \\24 rue Lhomond, 75005 Paris, France.}
\maketitle

\newcommand{\rhov}{\rho_{\rm v}}
\newcommand{\op}{\omega_{\perp}}
\newcommand{\rp}{r_{\perp}}
\newcommand{\oq}{\omega_{\rm QP}}
\newcommand{\oc}{\Omega_{\rm c}}
\newcommand{\oth}{\Omega_{\rm th}}
\newcommand{\ob}{\bar{\omega}_{\perp}}

\newcommand{\be}{\begin{equation}}
\newcommand{\ee}{\end{equation}}
\newcommand{\prl}{Phys.~Rev.~Lett.~}
\newcommand{\cs}{$\clubsuit$}
\newcommand{\fred}{$\spadesuit$}
\newcommand{\blackbox}{\vrule height 2mm width 2mm}

\begin{abstract}
Using  a focused laser beam we stir a Bose-Einstein condensate
confined in a magnetic trap. When the stirring frequency lies
near the transverse quadrupolar mode resonance we observe the
nucleation of vortices. When several vortices are nucleated,
they arrange themselves in regular Abrikosov arrays, and in
the limit of large quantum number the lattice structure
is shown to produce a quantum velocity field approaching that
for classical, rigid body rotation. Using a percussive
excitation of the condensate, we measure the angular momentum
of the condensate with vortices present and study the
nucleation band as a function of the stirring intensity
and geometry.  We find that with only quadratic terms in the
rotating perturbation the nucleation band is located around
the quadrupolar resonance and has a width that increases with
the strength of the stirring perturbation. However, when the
potential includes cubic terms, the nucleation band broadens
to include the hexapolar resonance as well. The results
presented here demonstrate that the nucleation of vortices
in the case of a harmonically trapped BEC is driven by the
resonant excitation of the rotating quadrupolar mode, or by
higher order rotating surface modes when the rotating
perturbation includes the corresponding terms.
\end{abstract}

\section{Introduction}
Superfluidity,  originally discovered and studied in the
context of superconductors and later in the system of
superfluid liquid Helium, is a hallmark property of
interacting quantum fluids and encompasses a whole class
of fundamental phenomena \cite{Lifshitz,Donnelly}.
With the achievement of Bose-Einstein condensation (BEC)
in laser-precooled atomic gases
\cite{Anderson95,Bradley957,Davis95,Fried98}, it became
possible to study these phenomena in an extremely dilute
quantum fluid, thus helping to bridge the gap between
theoretical studies, only tractable in dilute systems, and
experiments.

One striking consequence of superfluidity is the response
of a quantum fluid to a rotating perturbation.  In contrast
to a normal fluid, which at thermal equilibrium will rotate
like a solid body with the perturbation, a superfluid will not
circulate unless the frequency of the perturbation is larger
than some critical frequency, analogous to the critical
velocity \cite{Lifshitz}.  Moreover, when the superfluid does
circulate, it can only do so by forming vortices in which the
condensate density vanishes and for which the velocity field
flow evalutated around a closed contour is quantized.  Due
to their mutual repulsion these vortices can cristallize into
a regular lattice known in condensed matter as an Abrikosov
lattice \cite{Abrikosov57}.

The observation of superfluid phenomena in a dilute BEC has
been the subject of much experimental and theoretical work
during the last two years. A first class of study has been
performed at MIT and consists in measuring the energy deposited
in the condensate by a moving ``object" (a hole created by a
repulsive laser)\cite{Raman99}. The second class of study is
based on specific oscillation patterns of the condensate, in
particular the {\it scissors mode}. This was studied
theoretically by D.~Gu\'ery-Odelin and S.~Stringari \cite{dgo}
and investigated experimentally by the Oxford group
\cite{Marago00}. Finally another clue for superfluidity is
related to the behavior of vortices.  Using a
``phase printing method", the Boulder group has created a vortex
in a double component condensate with one component standing
still at the center of a magnetic trap and the other component
in quantized rotation around the first one \cite{Matthews99}.
In subsequent experiments this group also succeeded in emptying
the core of the double component vortex \cite{Anderson00}, and
showed that vortex rings can form as decay products of a
dark soliton \cite{Anderson00b}.

In our work in Paris, we have studied the response of a trapped,
single component condensate to a rotating perturbation created
by a stirring laser beam. When the stirring frequency lies near
the transverse quadrupolar mode resonance we observe the
nucleation of vortices \cite{Madison00}. We also find that
multiple vortices can be nucleated and that they spontaneously
arrange themselves into regular Abrikosov lattices containing up 
to 14 vortices for our experimental condition \cite{Madison00b}. 
A similar observation has been made quite recently
at MIT with a larger sodium condensate, in which large arrays 
of vortices (up to 150) have been nucleated \cite{Aboshaeer01}. 

We observe that in accordance with
the correspondence principle, the large quantum number limit
yields a quantum velocity field that approaches that for
classical, rigid body rotation.  Using a percussive excitation
of the condensate, we measure the angular momentum of the
condensate with vortices present and study the nucleation band
as a function of the stirring intensity and geometry
\cite{Chevy00}.  We find that with only quadratic terms in the
rotating perturbation the nucleation band is located at
the quadrupolar resonance and has a width that depends on the
strength of the stirring perturbation. However, when the potential
includes cubic terms, the nucleation band broadens to include
the hexapolar resonance as well. The results presented here
demonstrate that the nucleation of vortices in the case of a
harmonically trapped BEC is driven by the resonant excitation
of the rotating quadrupolar mode, or by higher order rotating
surface modes when the rotating perturbation includes the
corresponding terms.

Our experiment is a direct transposition to atomic gases of the
famous ``rotating bucket experiment" performed with liquid helium.
In  \S~\ref{He} we briefly recall some essential results obtained
with this experimental scheme for superfluid He.  We then turn
in \S~\ref{setup} to the description of our experimental setup.
The observation of single and multiple vortices is presented in
\S~\ref{vortex} where we also discuss the lattice structure.
Finally, in \S~\ref{nucleation} we discuss the measurement of
the angular momentum of the condensate as a function
of the stirring intensity and geometry.

\section{The rotating bucket experiment}
\label{He}

\begin{figure}[b]
\centerline{\epsfig{file=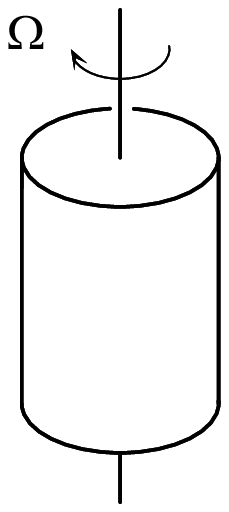,height=5.1cm,width=2.2cm}}
\caption{The rotating bucket experiment.}
\label{bucket}
\end{figure}

Consider a superfluid placed in a bucket rotating at angular
frequency $\Omega$ (fig.~\ref{bucket}).  If $\Omega$ is smaller
than a critical value $\oc$, the superfluid will not circulate, which is is a direct
manifestation of superfluidity. The existence of this critical angular
frequency is analogous to that of a
critical linear flow velocity below which the condensate
exhibits viscous free behaviour. Here the slow motion of the
rough walls of the bucket does not rotate the superfluid.
As shown by Landau, superfluidity is a direct consequence of
repulsive interactions which gives rise to a phonon-like
dispersion relation for the lowest lying excitations
\cite{Lifshitz}.

When $\Omega$ is increased beyond the critical frequency
$\oc$ the superfluid is set into motion.  As pointed out by
Onsager \cite{Onsager} and Feynman \cite{Feynman} the
corresponding velocity field is subject to very strong
constraints due to its quantum nature.
Consider the macroscopic wave function $\psi({\bf r})$
describing the state of the superfluid.  It can be written:
\begin{equation}
\psi({\bf r})=\sqrt{\rho({\bf r})}\, \exp{i\theta({\bf r})}\ ,
\end{equation}
where $\rho({\bf r})$ is the superfluid density.
The corresponding velocity field is given by:
\begin{equation}
{\bf v}({\bf r})=\frac{\hbar}{M}\nabla \theta({\bf r})\ .
\label{eq:vel}
\end{equation}
where $M$ is the mass of one atom.  From Eq.~(\ref{eq:vel})
it is clear that the circulation of the velocity along any
closed contour is quantized as a multiple of $h/M$:
\begin{equation}
\oint {\bf v}\cdot d{\bf r}=n\frac{h}{M}\qquad
\mbox{where }n\mbox{ is an integer.}
\label{quantifcirc}
\end{equation}
Just above $\oc$, the superfluid wave function has a singular
line or {\it vortex line}, along which the density is zero.
On any closed path going around this line the phase of the wave
function varies continuously from $0$ to $2\pi$.

It is worth emphazing that these vortices are universal
structures associated with a circulating quantum flow.  Besides
superfluid liquid helium, other large quantum systems such as
neutron stars and superconductors support quantized vortices.
In the latter case the rotation vector is induced
by an applied magnetic field which modifies the motion of the
charges, and the quantization of the circulation of the velocity
field results in magnetic flux quantization.

For a rotating frequency $\Omega$ notably larger than $\oc$,
several vortex lines can be generated, and they form a regular
triangular lattice.  Such regular lattices were orginally
predicted to occur in type-II superconductors by Abrikosov
\cite{Abrikosov57} and were subsequently observed experimentally
by imaging with an electron microscope small ferromagnetic
particles which when scattered on the surface of a
superconductor in a magnetic field accumulate at the flux line
exit points \cite{Trauble68,Sarma68}. Evidence of vortex lattice
arrangements in liquid helium was demonstrated by trapping
electrons at the core of each vortex and then accelerating the
electrons along the vortex lines to a phosphorus screen
\cite{Yarmchuk79}.

\section{The experimental setup}
\label{setup}

\begin{figure}[b]
\centerline{\epsfig{file=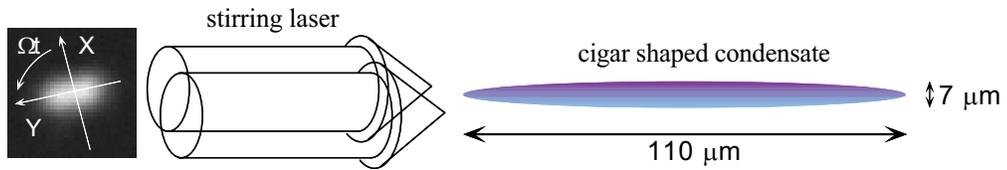,height=0.98in,width=5.21in}}
\caption{The cigar-shaped condensate is strirred using the
dipole potential created by a laser ``spoon". The laser waist
is 20~$\mu$m; its axis is toggled between two symmetric positions
about the trap axis, separated by 16~$\mu$m. The laser intensity
profile averaged after this toggling is displayed on the left of
the figure. The resulting anisotropic potential is then
rotated at the stirring frequency $\Omega$.}
\label{strirring}
\end{figure}

Our experiment is performed with a $^{87}$Rb gaseous condensate
(fig.~\ref{strirring}) confined in an Ioffe-Pritchard magnetic
trap.  The magnetic trap is an axisymmetric  harmonic potential:
\begin{equation}
U({\bf r})=\frac{1}{2}M\op^2(x^2+y^2) + \frac{1}{2}M\omega_z^2 z^2\ ,
\end{equation}
with a transverse frequency  $\op/2\pi$ varying from 90 to 225~Hz and
an axial frequency  $\omega_z/2\pi$ between 10~Hz and 12~Hz.  The
condensate is cigar-shaped, with a length of $\sim 110\;\mu$m
and a diameter of 7~$\mu$m for $2\times 10^5$ atoms
and a trapping frequency of 170~Hz.

The stirring of the condensate is provided by a
focused 500~$\mu$W laser beam
of wavelength $852$~nm and waist $w_0=20\;\mu$m, whose motion
is controlled using acousto-optic deflectors.
This laser propagates along the axis of the cigar, and it toggles
back and forth very rapidly between two symmetric positions $8\; \mu$m
from the center of the condensate. The toggling frequency is chosen
to be 100 kHz which is much larger than both $\omega_z$ and $\op$.
It creates for the atoms an average dipole potential which is
anisotropic
in the $xy$ plane (fig.~\ref{strirring}).  This potential contains
only even terms because of symmetry and the leading order term is:
\begin{equation}
\label{eq:potential}
\delta U({\bf r})=\frac{1}{2}M\op^2(\epsilon_X X^2+\epsilon_Y Y^2)
\end{equation}
where $\epsilon_X$ and $\epsilon_Y$ depend on the intensity, waist,
and the spacing of the stirring beams.  Using the acousto-optic
deflectors, the $XY$ axes are rotated at an angular frequency
$\Omega$ producing a rotating harmonic trap characterized by the
three trap frequencies $\omega_{X,Y}^2 = \op^2 (1+\epsilon_{X,Y})$
and $\omega_z$.  Using
\begin{equation}
X=x\cos (\Omega t)+y\sin(\Omega t)\qquad
Y=-x\sin(\Omega t)+y\cos(\Omega t)
\end{equation}
we can write the total potential in the rotating and in the lab
frame as:
\begin{eqnarray}
\left(U+\delta U \right)(\bf r) &=& \frac{1}{2}M
(\omega_X^2 X^2 +\omega_Y^2 Y^2)+\frac{1}{2}M\omega_z^2 z^2
\label{rotframe}
%\quad \mbox{rot. frame}
\\
&=& \frac{1}{2}M\bar \omega_\perp^2(x^2 +y^2)+\frac{1}{2}M\omega_z^2
z^2 \nonumber \\
&& +\frac{1}{2}M \,\epsilon \,\bar\omega_\perp^2
\left((x^2-y^2)\cos(2\Omega t)+2xy\sin(2\Omega t)  \right)
\label{labframe}
\end{eqnarray}
where we have set
\begin{equation}
\ob = \sqrt{(\omega_{X}^2+\omega_{Y}^2)/2}\qquad
\qquad
\epsilon = (\omega_{X}^{2} - \omega_{Y}^{2})/(\omega_{X}^{2} +
\omega_{Y}^{2})\ .
\end{equation}
This potential is stationary in the rotating frame (eq.
(\ref{rotframe})) and oscillates periodically at frequency
$2\Omega$ in the laboratory frame (eq. (\ref{labframe})).
The stirring frequency $\Omega$ is usually chosen in the
interval $(0,\op)$. At the upper value of this interval,
the centrifugal force equals the transverse restoring force of
the trap. A dynamical instability
of the center of mass motion of the condensate occurs
when $\Omega$ lies in the interval $[\omega_Y,\omega_X]$ (assuming
$\epsilon_Y<\epsilon_X$). The steepness of this
parametric resonance provides a very accurate measurement of the
trap frequencies $\omega_X$ and $\omega_Y$.

The experimental procedure begins with
the loading of $10^8$ $^{87}$Rb atoms into a magneto-optic trap from a
cold atomic jet produced by a separate magneto-optic trap
loading from a room temperature
vapor \cite{Wohlleben01}.
We then precool the atoms to 10~$\mu$K in an optical molasses and
transfer them into the magnetic trap.
The generation of a condensate in the pure magnetic potential
is provided by a radio-frequency (rf) evaporation ramp lasting 25~s.
The atomic cloud reaches the critical temperature
$T_{\rm c}\sim$~500~nK with an atom
number of $ \sim 2.5\,10^6$.  The evaporation is continued below
$T_{\rm c}$ to a temperature of or below 100~nK
at which point $3\;(\pm 0.7)\;10^5$ atoms are left in the condensate.
The rf frequency is then set 20~kHz
above $\nu_{\rm rf}^{\rm min}$, the rf frequency which corresponds
to the bottom of the magnetic potential.  This rf drive is kept
present
in order to hold the temperature approximately constant.
At this point, the stirring laser is switched on
and the condensate is allowed to evolve in the combined magnetic and
optical potential for a controlled duration.  Finally, the
stirring potential is extinguished adiabatically (in a time
long compared to $\op^{-1}$) and the
condensate density profile along the stirring axis
is measured to detect the presence of vortices.

We now address the question of the vortex visibility.
In a fluid of density $\rho$,
the radius of the vortex core is determined by the healing
length $\xi=(8\pi \rho a)^{-1/2}$, where $a$ is the scattering
length characterizing the two-body interactions in the
ultra-low temperature regime \cite{Lifshitz}.
For our experimental conditions, $\xi\sim 0.2\;\mu$m, which is too
small to
be observed optically with resonant light at $780$~nm.
Fortunately this size can be expanded
using a time-of-flight technique \cite{Castin96,Lundh98,Dalfovo99}.
When we release the atomic cloud from the magnetic trap and
let it expand for a duration $T$, the transverse dimensions of
the condensate and of the vortex core are increased by a factor
of $(1+\op^2T^2)^{1/2}\sim 35$ for $T=27$~ms and a transverse
trapping frequency of $\op/2\pi=170$~Hz.
The detection of the expanded condensate density profile
is then performed by imaging the absorption of a resonant laser beam
propagating along the $z$ axis.

\section{Single and multiple vortices}
\label{vortex}

\begin{figure}[t]
\centerline{\epsfig{file=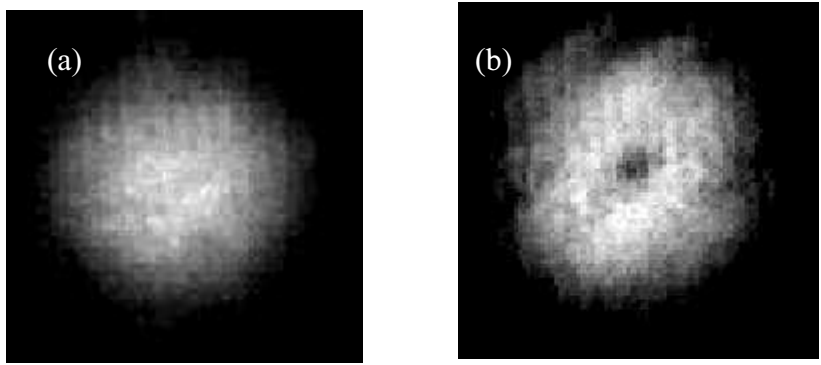,height=4.79cm,width=10.78cm}}
\caption{Density profile after a time-of-flight expansion
of a Bose-Einstein condensate stirred below (left) and above
(right) the threshold frequency for nucleating a vortex.
$\Omega/2\pi=145$~Hz and 152~Hz for the left and the right
column respectively. The number of atoms is $N=1.4\times 10^5$,
the temperature is below 80~nK and $\oth/2\pi= 147$~Hz. }
\label{vortex01}
\end{figure}

We now discuss the results of this experiment. When the stirring
frequency
is below a threshold frequency $\oth$ depending on the stirring
ellipticity,
no modification of the condensate is observed. Just above this
critical frequency
(within 1 or 2 Hz), a density dip appears at the center of the cloud,
with a
reduction of the optical thickness
at this location which reaches 50\% (fig.~\ref{vortex01}).

\begin{figure}[b]
\centerline{\epsfig{file=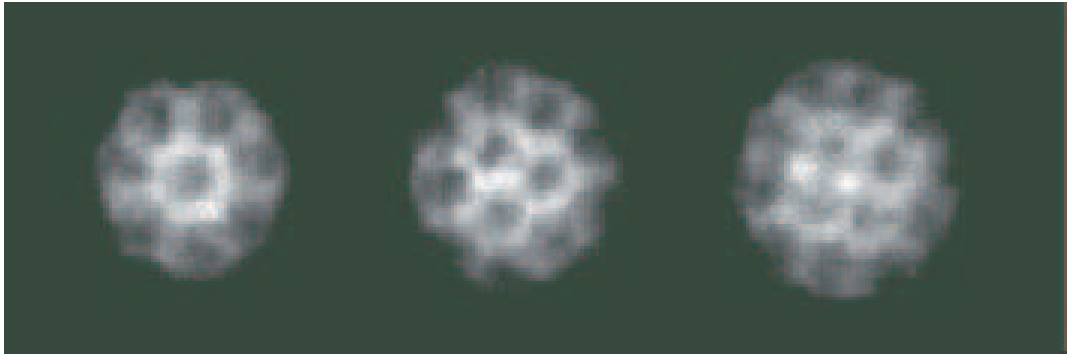,height=4.66cm,width=14.12cm}}
\caption{Arrays of vortices obtained in a magnetic trap with
$\ob/2\pi=103$~Hz and $\Omega/2\pi= 77$~Hz. For large number of vortices,
although the lowest energy configuration is a triangular lattice (middle image), the
square pattern is only slightly higher in energy and is sometimes
observed (right image).}
\label{vort81214}
\end{figure}

When we stir the condensate at a frequency higher than
$\oth$, more vortices are nucleated
(fig.~\ref{vort81214}).  The maximum number of vortices generated
in this experiment depends on the transverse oscillation frequency
$\ob$.
For a relatively tight trap ($\ob/2\pi=225$~Hz), we have observed
up to 4 vortices \cite{Madison00}.  When we used a less confining
trap
($\ob/2\pi \sim 100$~Hz), we obtained configurations where more
vortices
were present \cite{Madison00b}.  In
fig.~\ref{vort81214} we show images of vortex
lattices generated in a trap with $\ob/2\pi=103$~Hz,
where we obtained configurations
with up to 14 clearly visible vortices.

The existence of regular vortex lattices is a consequence
of the balance between the repulsive interaction between
two vortex lines and the restoring force that acts on a
vortex line centering it on the condensate (see
fig.~\ref{vortex01}).
Although the lowest energy configuration is a triangular lattice,
the square pattern is only slightly higher in energy and is sometimes
observed
(see fig.~\ref{vort81214}) \cite{Tkachenko66,Donnelly}.

In the large vortex number limit, the density of vortices
can be deduced from the {\em correspondence principle}.
In this limit, the coarse grain average (on a scale larger
than the distance between two vortices) of the quantum
velocity field should be the same as that for classical,
rigid-body rotation ${\bf v}={\bf \Omega} \times {\bf r}$
\cite{Feynman}.

In order to recover this linear variation of the velocity
field with ${\bf r}$, the distance from the rotation axis,
the surface density of vortices $\rho_v$ must be uniform.
The circulation of the velocity field
$\oint {\bf v}\cdot d{\bf r}$
on a circle of radius $R$ centered on the condensate
is then ${\cal N}(R)\; h/M$, where ${\cal N}(R)=\rhov \pi R^2$
is the number of vortices contained in the
circle (see eq. (\ref{quantifcirc})).
This is equivalent to the rigid body circulation
 $2\pi R^2 \Omega$ if
\be
\rhov=\frac{2M\Omega}{h} \ .
\ee

\begin{figure}[b]
\centerline{\epsfig{file=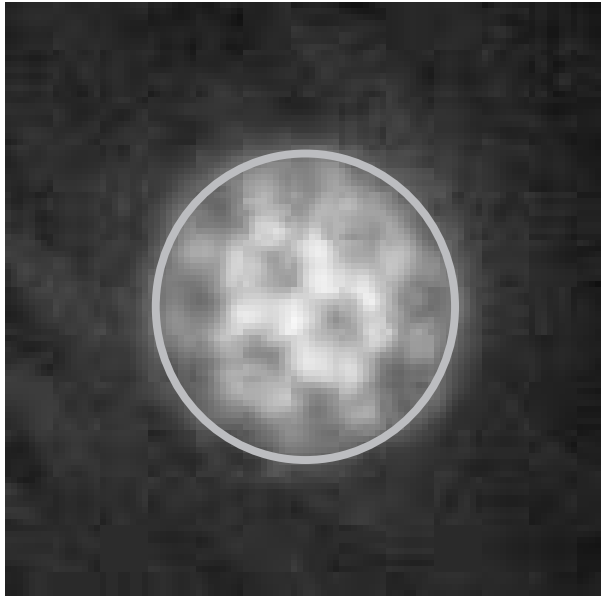,height=7.94cm,width=7.94cm}}
\caption{Array of ${\cal N}=12$ vortices obtained in a magnetic trap
with
$\ob/2\pi=103$~Hz and $\Omega/2\pi= 77$~Hz.
The circulation of the velocity field on the circle is given by
${\cal N}h/m$,
and is comparable with the one expected from rigid body rotation (see
text).}
\label{measurecirc}
\end{figure}

Consider for instance the vortex lattice shown in
fig.~\ref{measurecirc}, which contains ${\cal N}=12$ visible
vortices. The circle drawn on the edge of the condensate has a
radius 80 $\mu$m after the time-of-flight expansion, i.e.
$R=5\;\mu$m before expansion (using the expansion factor given
above $\sqrt{1+\op^2T^2}$). The stirring frequency in this
experiment was $\Omega/2\pi=77$~Hz, which yields the ratio
between the average velocity $\bar v$ on the circle and the
velocity $v_r$ corresponding rigid body rotation:
\begin{equation}
\frac{\bar v}{v_r}=\frac{{\cal N} \hbar}{M\Omega R^2} \simeq 0.7
\ .
\end{equation}
This shows that the coarse grain average velocity field of 
the condensate shown in fig.~\ref{measurecirc} is close 
that of a rigid body rotating at the stirring frequency $\Omega$.

\section{Vortex nucleation versus stirring intensity and geometry}
\label{nucleation}

After this qualitative observation of vortex nucleation and
lattice structure, we turn to a more quantitative observation
of vortex nucleation provided by the measurement of the angular
momentum per particle in a condensate with vortices.  For this
purpose we employ a theoretical result derived by F. Zambelli
and S. Stringari \cite{Zambelli98}. These authors have studied
the two transverse quadrupole modes of a cylindrically symmetric
condensate, corresponding respectively to excitations with angular
momentum $m=2$ and $m=-2$.  Because of symmetry for a condensate
at rest, these two modes have the same frequency. However, for
a condensate with a net circulation,  the degeneracy is lifted,
and the frequency difference between the two modes is related to
the average angular momentum per particle  $\langle L_z\rangle$
and to the transverse size of the condensate $\langle \rp^2\rangle$:
\begin{equation}
\omega_+-\omega_-=\frac{2\langle L_z\rangle}{M\langle \rp^2\rangle}
\label{Zamb}
\end{equation}

The measurement of the difference $\omega_+-\omega_-$ is
performed by looking at the transverse quadrupolar oscillation
of the condensate which is produced by a superposition of both
the $m=2$ and $m=-2$ excitations \cite{Chevy00,JILA}.
In the absence of vortices $\omega_+=\omega_-$ and
the oscillation occurs along fixed axes.
However, if $\omega_+$ and $\omega_-$ differ, the
axes of the quadrupolar oscillation precess,
and the precession frequency is $\dot \theta =
(\omega_+-\omega_-)/4$.

\begin{figure}[t]
\centerline{\epsfig{file=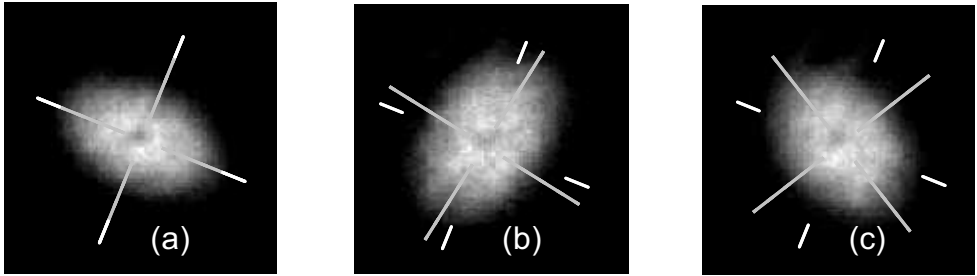,height=3.64cm,width=13.89cm}}
\caption{Transverse quadrupolar oscillations of a
condensate with $N\sim 3.7 \times 10^5$ atoms and
$\op/2\pi=171$~Hz. The stirring frequency is 120~Hz,
slightly above the vortex nucleation threshold
$\oth/2\pi=115$~Hz.  For a,b,c: the images were taken
$\tau=1,3,5$~ms after the excitation of the
quadrupolar oscillation.  The fixed axes indicate the
excitation basis and the rotating ones the
condensate axes. A single vortex is visible at the center of the
condensate. }
\label{precess}
\end{figure}

To study the angular momentum $\langle L_z\rangle$
of the condensate, we first stir it with the rotating laser
potential.  We then excite the transverse quadrupolar oscillation
using again the dipole potential created by the
stirring laser but now with a fixed basis ($x,y=X,Y$). This
potential is applied to
the atoms for a duration of 0.3~ms, which is short compared to the
quadrupolar oscillation
period $2\pi/\oq$ ($\oq=\sqrt2\, \op$).
We then let the atomic cloud oscillate freely in the pure magnetic
trap for
an adjustable period $\tau$, between 0 and 8~ms, after which we
perform the
time-of-flight + absorption imaging sequence. A typical result is
shown in fig.~\ref{precess}.
It displays three images taken 1, 3 and 5~ms after the quadrupolar
excitation of a condensate with a vortex present.
The precession of the quadrupolar axes is clearly visible and has an
angular velocity
$\dot \theta = 5.9$~degrees/ms. For this set of pictures, the
transverse trapping frequency
is 172~Hz
and  the {\it in situ} transverse size of the condensate
(inferred from the size measured after time-of-flight and the
expansion factor $(1+\op^2T^2)^{1/2}$)
is ${\langle \rp^2\rangle^{1/2}}=2.0\;\mu$m. From $\dot \theta$ and
$\langle \rp^2\rangle$
we deduce the value $L_z= 1.2\;(\pm 0.1)\; \hbar$ from the data shown
in fig.~\ref{precess}. A similar observation has also 
been made in Boulder \cite{Haljan01}.
This measurement is analogous to the experiment performed
with superfluid liquid helium by Vinen in which he detected a single
quantum of circulation
in rotating He II by measuring the frequencies of two opposing
circular vibrational modes
of a thin wire placed at the center of the rotating fluid
\cite{Vinen}.

\begin{figure}[t]
\centerline{\epsfig{file=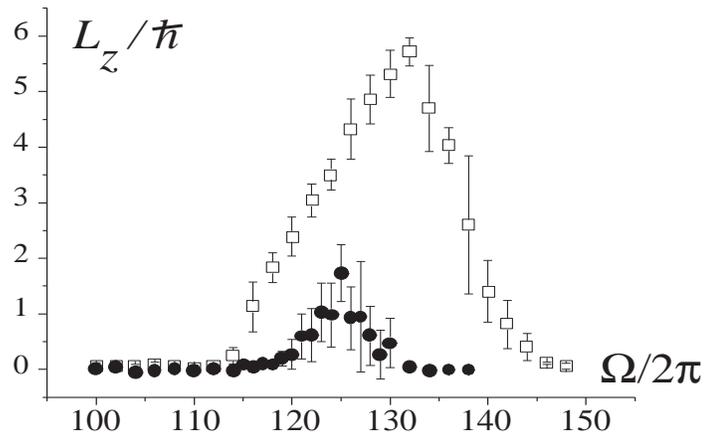,height=2.27in,width=3.64in}}
\caption{ Angular momentum per particle of the condensate
$L_z$ deduced from Eq.~(\ref{Zamb}) as a function of the
stirring frequency $\Omega$
and ellipticity $\epsilon$.
($\bullet$: $\epsilon=0.010\pm 0.002$; $\Box$: $\epsilon=0.019 \pm 0.004$).
$\op/2\pi=172$~Hz and $N=2.5\times10^5$ atoms.}
\label{Lz}
\end{figure}

We have repeated this angular momentum measurement for various
stirring
frequencies and ellipticities, and the results for two intensities
are shown in fig.~\ref{Lz}.  Both plots display a region where vortices
are nucleated, and the more intense the
stirring anisotropy the wider is this region. For very small stirring
intensities,
this region has a  width of only a few Hz, and it is centered on a stirring frequency
$2\pi\times 125$~Hz. This corresponds to the situation where the
time-dependent perturbation in the lab frame oscillating at frequency
$2 \Omega$ (see Eq. (\ref{labframe})) is resonant with the quadrupole
mode at frequency $\oq$. In other words, the resonance observed
for very low stirring intensities occurs when
$\Omega$ is close to the quadrupolar rotational resonance
$\oq/2$, which is equal to $\sqrt{2}\,\ob/2\simeq
2\pi\times 122$~Hz
in the Thomas Fermi limit \cite{qpexplication}. This coincidence and
the observation of very strong
BEC ellipticities during the first tens of milliseconds of stirring
indicates
that the mechanism for vortex nucleation involves the resonant
excitation of
the rotating quadrupole mode \cite{Dalfovo00}.  The details of
how the rotating quadrupole mode leads to the nucleation
of vortices after its excitation by the rotating potential
is beyond the scope of this paper, and they are discussed in
\cite{Madison01}.

We note here that the percussive measurement
of $L_z$ is performed 200~ms after the stirring anisotropy
is switched off.  This delay is (i) short compared
to the vortex lifetime \cite{Madison00} and (ii)
long compared to the measured relaxation time ($\sim 25$~ms)
of the aforementioned ellipticity of the condensate excited by the
stirring.
Therefore a nonzero value of $L_z$ is a clear signature of vortex
nucleation.

The role of the resonant excitation of the transverse quadrupole mode
in vortex nucleation suggests the possibility that vortices might
be created by exciting transverse collective modes of higher angular
momenta, $m$, at their resonant rotational frequencies
$\Omega_{m}\simeq \ob/\sqrt{m}$
\cite{qpexplication,Dalfovo00,Onofrio00,Aboshaeer01}.
In order to investigate
this possibility, the stirring potential must include non-negligible
terms of higher order in $X$ and $Y$.  Although the two-beam
stirring potential whose dominant term is given in Eq.~\ref{eq:potential} contains
in principle
all even orders, their magnitude is too small in practice to be relevant for
$m\ge4$.
Using a single-beam stirring potential (i.e. suppressing the toggling
motion)
the odd terms are no longer canceled.
To leading order, the resulting potential then contains the same
quadratic term as before (Eq.~\ref{eq:potential}) as well as a
linear and cubic term.  Although the linear term can directly
excite the center of mass motion of the condensate, a time dependent
displacement of the center of a harmonic trap produces
only a movement of the center of mass of the wavefunction and no
deformation
even in the presence of interactions \cite{Dobson94}.

\begin{figure}[t]
\centerline{\epsfig{file=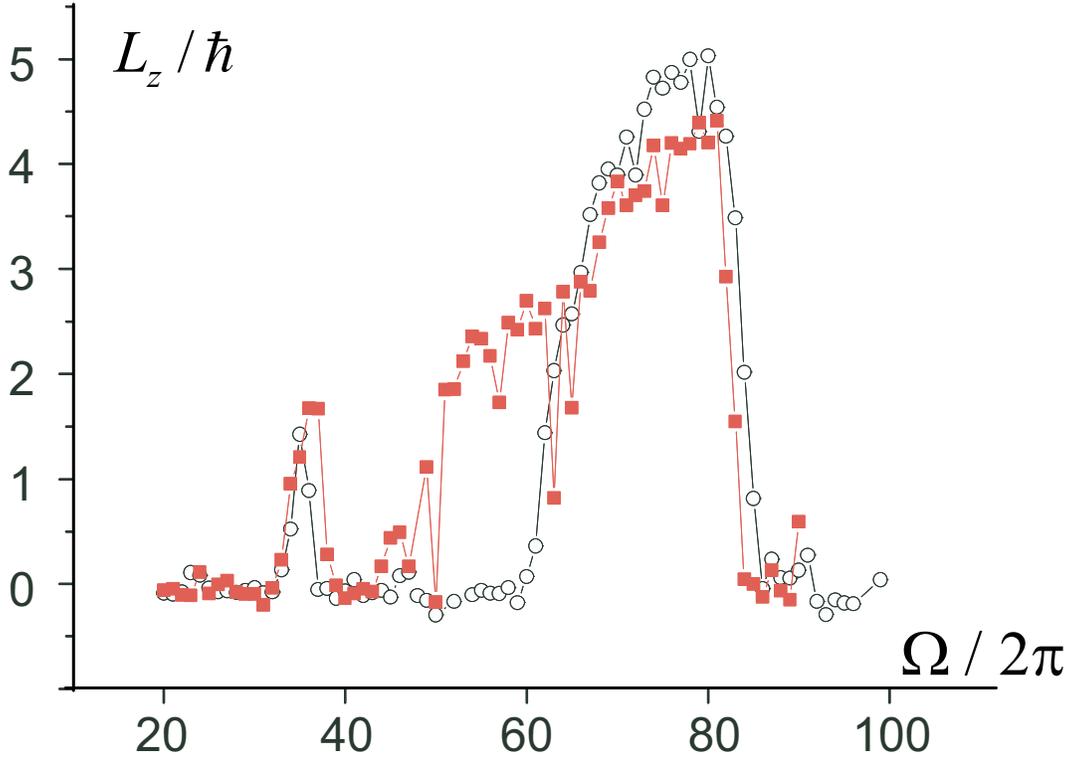,height=4.37in,width=6.02in}}
\caption{Angular momentum per particle
versus stirring frequency for two different
geometries of stirring potentials ($\ob/2\pi=100$~Hz).
The nucleation band with the usual two-spot configuration is
shown by the hollow circles $\circ$ and the results
with a single spot are shown by the filled squares
\blackbox.  In the latter case, because the potential
contains terms of order three in $X$ and $Y$,
the nucleation range increases due to the hexapolar
resonance located at 58~Hz. The nucleation peak near
35~Hz which occurs for both stirring potentials
corresponds to a rotating quadrupolar resonance excited by
an (unintentional) intensity modulation of the stirring
potential at twice the stirring frequency (see text).}
\label{monobipode}
\end{figure}

The presence of the cubic term, however, allows for the
resonant excitation of the rotating hexapole mode ($m=3$)
at a frequency of $\ob/\sqrt{3} \simeq 0.58\, \ob$.
fig.~\ref{monobipode} shows the results of
an angular momentum measurement for various stirring frequencies with
both a two beam and single beam stirring scheme where
$\ob/2\pi = 100$~Hz.  In the case of the single
stirring beam, the nucleation band includes the
quadrupolar resonance centered at 71~Hz and extends beyond
this down to 45~Hz including the hexapole resonance
at 58~Hz.

The vortex nucleation peak near 35~Hz present
on the two curves of fig.~\ref{monobipode}
is due to the difference of the diffraction efficiencies of the
two acousto-optic deflectors controlling the the stirring beam
position.
This slight difference of efficiency creates a small modulation
of the intensity of the stirring potential at twice the stirring
frequency.
This intensity modulation therefore modulates
the transverse trap frequency at 70~Hz,
and this modulation can excite parametrically the quadrupolar
resonance.
We indeed observe in this case the excitation of the rotating
quadrupole mode and its subsequent decay into vortex states as
before.
We emphasize that nucleation at this low frequency
is not relevant in the usual study of vortex nucleation,
since in the rotating frame the potential is no longer
stationary: actually it is not stationary in any frame!

\section{Conclusions}

In conclusion we report on a rotating bucket experiment
performed with a gaseous condensate.  We observe
(i) that there exists a frequency band in which vortices
can be nucleated; (ii) that for a quadratic rotating perturbation
this band is centered on a critical frequency $\oc$
identified with the rotating quadrupole mode;
(iii) that the width of this band can be significantly
increased by changing the stirring beam potential
to excite surface excitations of higher order, and
(iv) that multiple vortices can be nucleated and will arrange
themselves into regular Abrikosov lattices for which,
in the large quantum number limit, the velocity field
approaches that for classical, rigid body rotation.
The results presented here demonstrate that the nucleation
of vortices in the case of a harmonically trapped BEC is
related to the resonant (dynamical) excitation of the
rotating quadrupole mode and its subsequent decay into a vortex state.
This nucleation mechanism is different from the
scenarii expected from previous theoretical work which
concentrated on the study of thermodynamical instabilities.
Indeed one predicts in this case a critical
nucleation frequency much lower than what is observed, and
 a nucleation region extending to $\ob$
\cite{Baym96,Stringari96,Sinha97,Lundh97,Dalfovo97,Fetter98,Feder99,Castin99,Isoshima99,Garcia99,Fetter00,Feder00}.

The dynamical instability of the condensate in presence of a small rotating
anisotropy has been studied
by numerically integrating the zero temperature Gross-Pitaevski
equation
and found to produce states with nonzero angular momentum
\cite{Feder01,Yvan00}.
We have recently studied this issue experimentally and clarified
the excitation and decay routes of the rotating quadrupole mode
which lead to vortex nucleation \cite{Madison01}.
We now plan to pursue a detailed investigation of the
possible bending of vortex lines \cite{Garcia00} and of 
the
decay of vortices, which is relevant to the physics
of rotating neutron stars \cite{Fedichev99,Fedichev00}.

\newpage

\end{document}